# Reconstructed black hole solutions in the scalar-tensor theory with nonminimal coupling

K. K. Ernazarov

*Institute of Gravitation and Cosmology,*
*Peoples' Friendship University of Russia (RUDN University),*
*6 Miklukho-Maklaya Street, Moscow, 117198, Russian Federation,*

**Abstract**

We consider the scalar-tensor theory witn non-minimal coupling in the Jordan frame. The action of the model contains a potential term $U(\varphi)$, a coupling function $f(\varphi)$. We explore a reconstruction procedure for a generic static spherically symmetric metric written in the Buchdahl parametrization: $ds^2 = (A(u))^{-1} du^2 - A(u)dt^2 + C(u)d\Omega^2$, with given $A(u) > 0$ and $C(u) > 0$. The procedure gives the relations for $U(\varphi(u))$, $f(\varphi(u))$ and $d\varphi/du$, which lead to exact solutions to equations of motion with a given metric. A key role in this approach is played by the solutions to a first order linear differential equation for the function $f(\varphi(u))$. The formalism is illustrated by two examples when: a) the Reissner-Nordström-(Anti-)de Sitter metric and b) the Bocharova-Bronnikov-Melnikov-Bekenstein-(Anti)de-Sitter metric are chosen as a starting point.

## 1 Introduction

For over a century, General Relativity (GR) has stood as the cornerstone of our understanding of gravitation, describing the force as the curvature of spacetime caused by mass and energy. Despite its remarkable success, the pursuit of a more complete theory of gravity has never ceased. Among the myriad of alternatives proposed, the scalar-tensor theory of gravitation stands out as one of the most enduring, well-studied, and physically motivated extensions of Einstein's work . By introducing an additional scalar field alongside the metric tensor field of GR, these theories offer a richer, more flexible framework to explore phenomena ranging from the varying constants of nature to the mysterious accelerating expansion of the Universe.

The core innovation of scalar-tensor theory is the introduction of a new degree of freedom: a scalar field, often denoted as $\phi$. While GR's action depends only on the Ricci scalar $R$ (which encodes spacetime curvature), the action for a scalar-tensor theory includes both $R$ and the scalar field $\phi$, along with its kinetic energy and potential energy. The general action can be written in a form often associated with the Bergmann-Wagoner class of theories [1], [2]:

$$S = \int d^4 z \, |g|^{\frac{1}{2}} \left( \frac{f(\phi)}{2\kappa^2} R(g) - \frac{1}{2}\omega(\phi) g^{MN} \partial_M \phi \partial_N \phi - U(\phi) \right) + S_m[\psi, g_{MN}], \quad (1.1)$$

In this formulation, $f(\phi)$ is a function that couples the scalar field directly to curvature, signifying a nonminimal coupling. This is the hallmark of a scalar-tensor theory. The function $\omega(\phi)$ acts as a "Brans-Dicke coupling parameter", and $U(\phi)$ is a potential that can give the scalar field a mass or drive cosmic acceleration. The final term, $S_m$ represents the action for all matter fields ($\psi$), which couple universally only to the metric $g_{MN}$. This universal coupling is crucial because it ensures the Weak Equivalence Principle is satisfied - all test bodies fall at the same rate in an external gravitational field, a fact that has been experimentally verified to high precision.

The Brans-Dicke theory is the simplest case, where $f(\phi) = \phi$ and $\omega(\phi)$ is a constant, with no potential. In this model, the inverse of the scalar field acts as an effective gravitational constant ($G_{eff} \sim \frac{1}{\phi}$), making the variability of $G$ a central prediction.

A rich and sometimes confusing aspect of scalar-tensor theory is the use of conformal transformations. Mathematically, one can perform a transformation on the metric ($g_{MN} \to \tilde{g}_{MN} = \Omega^2(\phi) g_{MN}$) to rewrite the action in a different form. In the so-called *Jordan frame*, the scalar field is nonminimally coupled to curvature, and matter follows geodesics of the metric $g_{MN}$. In the transformed or *Einstein frame*, the gravitational part of the action looks exactly like that of General Relativity, but the scalar field becomes minimally coupled and now interacts directly with matter. A long-standing debate concerns which frame is "physical". While classical predictions are often frame-independent, the debate highlights the profound differences between scalar-tensor theories and GR, particularly regarding the violation of the Strong Equivalence Principle. This principle violation means that the motion of massive, self-gravitating bodies (like neutron stars) can depend on their internal composition, a key target for experimental tests.

In this scientific research area, we can identify the main scalar-tensor gravity theories with nonminimal coupling, that emerge from the Bergmann-Wagoner class of theories. They are:

- Barker's theory [3], in which the effective gravitational constant is really a constant:

$$f(\phi) = \phi, \quad \omega(\phi) = \frac{4 - 3\phi}{2\phi(\phi - 1)}. \tag{1.2}$$

Barker's scalar-tensor theory is a significant and conceptually clean branch of alternative gravity theories. Defined by the condition of a constant local gravitational constant $G$ (which fixes the coupling function $\omega(\phi)$), it serves as a valuable theoretical model for exploring the consequences of scalar-tensor gravity while automatically satisfying many local experimental constraints on gravitational constant's variation. Its primary value lies in its ability to generate exact cosmological and astrophysical solutions, offering a laboratory to study the role of scalar fields in the universe's dynamics, from the initial singularity to the formation of black holes and the nature of gravity itself.

- Schwinger's theory [4, 5]:

$$f(\phi) = \phi, \quad \omega(\phi) = \frac{K - 3\phi}{2\phi^2}, \quad K = \text{const} > 0. \tag{1.3}$$

Schwinger's scalar-tensor theory is a significant and pragmatic branch of alternative gravity theories. It serves as a simpler, analytically tractable model within the Nordtvedt class. Its primary value lies in its ability to generate exact cosmological solutions, offering a theoretical laboratory to study the role of scalar fields in the early universe's dynamics, the nature of dark energy, and the behavior of gravity beyond Einstein's theory.

This paper is structured as follows: Section 2 provides a brief introduction to the scalar-tensor gravity theory with nonminimal coupling, focusing on the reconstruction method. Since a more detailed analysis of this method is presented in our previous work [12]. As it is obvious from (2.29), the master equation is an homogeneous first-order differential equation. Section 3 is dedicated to the application of the reconstruction method results obtained in Section 2 to the the Reissner-Nordström-(Anti-)de Sitter metric and Bocharova-Bronnikov-Melnikov-Bekenstein-(Anti)de-Sitter metric. A detailed analysis of the derivation of Lagrangian (2.6) is shown in the Appendix. The paper concludes with a discussion of the results and final remarks.

## 2 The reconstructed scalar-tensor theory with nonminimal coupling

The scalar-tensor theory often considered as GR with a nonminimally (conformally or non-conformally) coupled scalar field:

$$f(\phi) = 1 - \xi\phi^2, \quad \omega(\phi) = 1, \quad \xi = \text{const.} \tag{2.1}$$

This theory splits into four cases with different expressions for $\psi(\phi)$), was discussed on [6]: (a) $\xi = 1/6$ (conformal coupling), (b) $0 < \xi < 1/6$, (c) $\xi > 1/6$, and (d) $\xi < 0$.

In this paper, we discuss the solutions of scalar-tensor theory with nonminimal coupling for an arbitrary function $f(\phi)$ in case of $\omega(\phi) = 1$. The action in Jordan frame has following form:

$$S = \int d^4z \, |g|^{\frac{1}{2}} \left( \frac{f(\phi)}{2\kappa^2} R(g) - \frac{1}{2} g^{MN} \partial_M \phi \partial_N \phi - U(\phi) \right), \tag{2.2}$$

where $\kappa^2 = 8\pi \frac{G_N}{c^4}$ ($G_N$ is Newton's gravitational constant, $c$ is speed of light), $\phi$ is scalar field, $g_{MN}dz^M dz^N$ is 4d metric of signature $(-,+,+,+)$, $R[g]$ is scalar curvature, $U(\phi)$ is potential, $f(\varphi)$ and $\omega(\phi)$ are arbitrary functions ($f(\phi) > 0$ describing a nonminimal coupling between the scalar field $\phi$ and the curvature).

We study the spherically-symmetric solutions with the metric

$$ds^2 = g_{MN}(z)dz^M dz^N = e^{2\gamma(u)}du^2 - e^{2\alpha(u)}dt^2 + e^{2\beta(u)}d\Omega^2 \tag{2.3}$$

defined on the manifold

$$M = \mathbb{R} \times \mathbb{R}_* \times S^2. \tag{2.4}$$

Here $\mathbb{R}_* = (r_h, +\infty)$, where $r_h$ is the Black hole horizon, and $S^2$ is 2-dimensional sphere with the metric $d\Omega^2 = d\theta^2 + sin^2\theta d\varphi^2$, where $0 < \theta < \pi$ and $0 < \varphi < 2\pi$.

By substitution the metric (2.3) into the action we obtain

$$S = 4\pi \int du \left( L + \frac{dF_*}{du} \right), \tag{2.5}$$

where the Lagrangian $L$ reads

$$L = \frac{f(\phi)}{\kappa^2}\left(e^{\alpha-\gamma+2\beta}\dot{\beta}\left(\dot{\beta}+2\dot{\alpha}\right)+e^{\alpha+\gamma}\right)+\frac{1}{\kappa^2}\dot{\phi}\frac{df}{d\phi}\left(\dot{\alpha}+2\dot{\beta}\right)e^{\alpha-\gamma+2\beta}-$$
$$\frac{1}{2}e^{\alpha-\gamma+2\beta}\dot{\phi}^2-e^{\alpha+\gamma+2\beta}U(\phi), \tag{2.6}$$

and the total derivative term is irrelevant for our consideration. Here and in what follows we denote $\dot{x} = \frac{dx}{du}$. The relation (2.5) is derived in Appendix, where an explicit relation for the $F_*$ term is given.

The equations of motion derived from the action (2.2) using the metric (2.3) are equivalent to the Lagrange equations obtained from the Lagrangian (2.6), and take the following form

$$\frac{\partial L}{\partial \gamma} = \frac{f}{\kappa^2}\left(-e^{\alpha-\gamma+2\beta}\dot{\beta}\left(\dot{\beta}+2\dot{\alpha}\right)+e^{\alpha+\gamma}\right)-\frac{1}{\kappa^2}\dot{\phi}\frac{df}{d\phi}e^{\alpha-\gamma+2\beta}\dot{\beta}\left(\dot{\alpha}+2\dot{\beta}\right)$$
$$+\frac{1}{2}e^{\alpha-\gamma+2\beta}\dot{\phi}^2-e^{\alpha+\gamma+2\beta}U(\phi)=0, \tag{2.7}$$

$$\frac{d}{du}\left(\frac{\partial L}{\partial \dot{\alpha}}\right)-\frac{\partial L}{\partial \alpha} = \frac{1}{\kappa^2}\frac{d}{du}\left(2fe^{\alpha-\gamma+2\beta}\dot{\beta}+\dot{\phi}\frac{df}{d\phi}e^{\alpha-\gamma+2\beta}\right)$$
$$-\frac{f}{\kappa^2}\left(e^{\alpha-\gamma+2\beta}\dot{\beta}\left(\dot{\beta}+2\dot{\alpha}\right)+e^{\alpha+\gamma}\right)+\frac{1}{\kappa^2}\dot{\phi}\frac{df}{d\phi}\left(e^{\alpha-\gamma+2\beta}\dot{\beta}\left(\dot{\alpha}+2\dot{\beta}\right)\right)- \tag{2.8}$$
$$\frac{1}{2}e^{\alpha-\gamma+2\beta}\dot{\phi}^2-e^{\alpha+\gamma+2\beta}U(\phi)=0,$$

$$\frac{d}{du}\left(\frac{\partial L}{\partial \dot{\beta}}\right)-\frac{\partial L}{\partial \beta} = \frac{2}{\kappa^2}\frac{d}{du}\left(fe^{\alpha-\gamma+2\beta}\left(\dot{\beta}+\dot{\alpha}\right)+\dot{\phi}\frac{df}{d\phi}e^{\alpha-\gamma+2\beta}\right)-$$
$$\frac{2}{\kappa^2}\left(fe^{\alpha-\gamma+2\beta}\dot{\beta}\left(\dot{\beta}+2\dot{\alpha}\right)+\dot{\phi}\frac{df}{d\phi}e^{\alpha-\gamma+2\beta}\left(\dot{\alpha}+2\dot{\beta}\right)\right)- \tag{2.9}$$
$$e^{\alpha-\gamma+2\beta}\dot{\phi}^2-2e^{\alpha+\gamma+2\beta}U(\phi)=0,$$

and

$$\frac{d}{du}\left(\frac{\partial L}{\partial \dot{\varphi}}\right)-\frac{\partial L}{\partial \varphi} = \frac{d}{du}\left(\frac{1}{\kappa^2}\frac{df}{d\phi}\left(\dot{\alpha}+2\dot{\beta}\right)e^{\alpha-\gamma+2\beta}-e^{\alpha-\gamma+2\beta}\dot{\phi}\right)$$
$$-\left(\frac{1}{\kappa^2}\dot{\phi}\frac{d^2f}{d\phi^2}\left(\dot{\alpha}+2\dot{\beta}\right)e^{\alpha-\gamma+2\beta}-e^{\alpha+\gamma+2\beta}\frac{dU}{d\varphi}\right)=0. \tag{2.10}$$

Now, we use (without loss of generality) the Buchdahl radial gauge obeying $\alpha = -\gamma$. For the metric (2.3) we obtain

$$ds^2 = (A(u))^{-1}du^2 - A(u)dt^2 + C(u)d\Omega^2, \tag{2.11}$$

where
$$e^{2\gamma(u)} = (A(u))^{-1}, \quad e^{2\alpha(u)} = A(u) > 0, \quad e^{2\beta(u)} = C(u) > 0. \tag{2.12}$$

In what follows we use the identities
$$\dot{\alpha} = \frac{\dot{A}}{2A}, \quad \dot{\beta} = \frac{\dot{C}}{2C}. \tag{2.13}$$

We put (without loss of generality) $\kappa^2 = 1$. We also denote
$$f(\phi(u)) = f, \quad U(\phi(u)) = U \tag{2.14}$$

and hence
$$\frac{d}{du}f = \frac{df}{d\phi}\frac{d\phi}{du} \iff \dot{f} = \frac{df}{d\phi}\dot{\phi}, \tag{2.15}$$

$$\frac{d}{du}U = \frac{dU}{d\phi}\frac{d\phi}{du} \iff \dot{U} = \frac{dU}{d\phi}\dot{\phi}. \tag{2.16}$$

Multiplying (2.7) by $(-2)$ and using relations (2.12), (2.13), (2.15) we get
$$\left(C\dot{A} + 2A\dot{C}\right)\dot{f} + 2\left(AK + \frac{\dot{C}\dot{A}}{2} - 1\right)f - CA\dot{\phi}^2 + 2CU = 0, \tag{2.17}$$

where here and in what follows we use the notation
$$K \equiv \left(\frac{\dot{C}}{2C}\right)^2 C. \tag{2.18}$$

Multiplying (2.8) by 2 and using relations (2.12), (2.13), (2.15), (2.18) we get
$$2AC\ddot{f} + \left(2A\dot{C} + C\dot{A}\right)\dot{f} + 2\left(\frac{\dot{C}\dot{A}}{2} - AK + A\ddot{C} - 1\right)f + CA\dot{\phi}^2 + 2CU = 0. \tag{2.19}$$

Analogously, using (2.12), (2.13), (2.15), (2.18) we rewrite equation (2.9) as follows
$$2AC\ddot{f} + \left(A\dot{C} + 2C\dot{A}\right)\dot{f} + \left(\dot{C}\dot{A} + A\ddot{C} + C\ddot{A} - 2AK\right)f + CA\dot{\phi}^2 + 2CU = 0. \tag{2.20}$$

Now, multiplying equation (2.10) by $2\dot{\phi}$ we obtain
$$\left(2A\ddot{C} + C\ddot{A} + 3\dot{A}\dot{C}\right)\dot{f} - 2\left(C\dot{A} + A\dot{C}\right)\dot{\phi}^2 - 2AC\dot{\phi}\ddot{\phi} -$$
$$\left(2A\dot{C} + C\dot{A}\right)\left(\ddot{f} - \frac{\ddot{\phi}}{\dot{\phi}}\dot{f}\right) + 2AC\dot{U} = 0. \tag{2.21}$$

Here we put the following restriction
$$\dot{\phi} \neq 0 \quad \text{for} \quad u \in (u_-, u_+), \tag{2.22}$$

where interval $(u_-, u_+)$ is belongs to $\mathbb{R}_* = (u_-, +\infty)$. Then the relations (2.21) and (2.10) are equivalent in this interval.

By adding equations (2.19) and (2.17) and dividing the result by $4C$ we get the relation for the function $U = U(\phi(u))$

$$U = -\frac{1}{2C}\left[CA\ddot{f} + \left(2A\dot{C} + \dot{A}C\right)\dot{f} - \left(2 - \dot{C}\dot{A} - A\ddot{C}\right)f\right]. \tag{2.23}$$

The relation (2.23) may be written as

$$U = E_U \ddot{f} + F_U \dot{f} + G_U f, \tag{2.24}$$

where

$$E_U = -\frac{A}{2}, \tag{2.25}$$

$$F_U = -\frac{1}{2C}\left(2A\dot{C} + \dot{A}C\right), \tag{2.26}$$

$$G_U = \frac{1}{2C}\left(2 - \dot{C}\dot{A} - A\ddot{C}\right). \tag{2.27}$$

Subtracting (2.17) from (2.19) and dividing the result by $2AC$, we obtain the relation for $\dot{\varphi}$

$$h(u) = \dot{\phi}^2 = \frac{1}{C}\left(2K - \ddot{C}\right)f - \ddot{f}. \tag{2.28}$$

Subtracting (2.19) from (2.20), we get the master equation for the function $f = f(\varphi(u))$

$$F\dot{f} + Gf = 0, \tag{2.29}$$

where

$$F = \left(\dot{A}C - A\dot{C}\right), \tag{2.30}$$

$$G = \left(2 + \ddot{A}C - A\ddot{C}\right). \tag{2.31}$$

The solution to differential equation (2.29) can be readily obtained by using standard methods. This solution reads

$$f = C_0 \exp\left(-\int \frac{G}{F} du\right), \tag{2.32}$$

where $u \in (u_-, u_+)$, $C_0$ is an arbitrary constant.

# 3 Examples

Here we consider the application of the results obtained in Section 2 to the Reissner-Nordström-(Anti-)de Sitter- and Bocharova-Bronnikov-Melnikov-Bekenstein-(Anti)de-Sitter metrics. The solutions on the base of scalar-tensor gravity theory in these metrics are often discussed in various scientific papers before [7] - [9] etc. This section presents more original results, that we have obtained based on the reconstructed scalar-tensor theory with nonminimal coupling.

## 3.1 Reissner-Nordström-(Anti-)de Sitter metric

Here, we present reconstruction examples to test and validate the reconstruction scheme under consideration. We apply the reconstructed scalar-tensor theory to the metric of the Reissner-Nordström-(Anti-)de Sitter (RN-(A)dS) black hole solution:

$$ds^2 = (A(u))^{-1} du^2 - A(u) dt^2 + C(u) d\Omega^2, \tag{3.1}$$

where

$$A(u) = \left(1 - \frac{2\mu}{u}\right) + \frac{Q^2}{u^2} - \frac{\lambda}{3} u^2,$$
$$C(u) = u^2, \tag{3.2}$$

In the given metric, for the master equations (2.29) of the functions F(u) and G(u), defined in (2.30) and (2.31), are expressed, respectively, as

$$F = \frac{2}{u}\left(3\mu u - 2Q^2 - u^2\right), \tag{3.3}$$

$$G = \left(\frac{2Q}{u}\right)^2. \tag{3.4}$$

Solving master equation $F\dot{f} + Gf = 0$ we obtain

$$f(\phi(u)) = C_0 \cdot \exp\left(-\int \frac{G}{F} du\right) = \frac{C_0 u}{\sqrt{3\mu u - 2Q^2 - u^2}} \exp\left(-\frac{\arctan\left(\frac{3\mu - 2u}{\sqrt{8Q^2 - 9\mu^2}}\right)}{\sqrt{\frac{8Q^2}{9\mu^2} - 1}}\right) > 0 \tag{3.5}$$

and

$$\dot{f} = -\frac{2C_0 Q^2}{(3\mu u - 2Q^2 - u^2)^{\frac{3}{2}}} \exp\left(-\frac{\arctan\left(\frac{3\mu - 2u}{\sqrt{8Q^2 - 9\mu^2}}\right)}{\sqrt{\frac{8Q^2}{9\mu^2} - 1}}\right),$$

$$\ddot{f} = \frac{3C_0 Q^2 (3\mu u - 2Q - u)(u - 2\mu)^2}{(3\mu u - 2Q^2 - u^2)^{\frac{7}{2}}} \exp\left(-\frac{\arctan\left(\frac{3\mu - 2u}{\sqrt{8Q^2 - 9\mu^2}}\right)}{\sqrt{\frac{8Q^2}{9\mu^2} - 1}}\right). \tag{3.6}$$

Here $C_0$ is an arbitrary constant. The function $f(\phi(u))$ must always be positive, so an arbitrary constant $C_0$ has only positive value.

In this case, for the scalar potential $U(u)$ from expressions (2.25), (2.26), (2.27) we obtain the following relationships:

$$E_U = -\left(\frac{1}{2} - \frac{\mu}{u}\right) - \frac{Q^2}{2u^2} + \frac{\Lambda}{6}u^2 \tag{3.7}$$

$$F_U = \Lambda u - \frac{2}{u} + \frac{3\mu}{u^2} - \frac{Q^2}{u^3}, \tag{3.8}$$

$$G_U = \Lambda + \frac{Q^2}{u^4} \tag{3.9}$$

and hence using (2.23) and (3.6) we get the following relation for potential function

$$U = \frac{C_0\left(3\Lambda\left(\frac{u^6}{3} - 2\mu u^5 + \left(3\mu^2 + \frac{5Q^2}{3}\right)u^4 + \frac{8Q^2u^2(Q^2-2\mu u)}{3}\right) + 3Q^2(\mu - Q)(\mu + Q)\right)}{(3\mu u - 2Q^2 - u^2)^{\frac{5}{2}}u} \times$$

$$\exp\left(-\frac{\arctan\left(\frac{3\mu - 2u}{\sqrt{8Q^2 - 9\mu^2}}\right)}{\sqrt{\frac{8Q^2}{9\mu^2} - 1}}\right). \tag{3.10}$$

From relation (2.28) we get

$$\dot{\phi}^2 = h(u), \tag{3.11}$$

$$h(u) = \frac{6C_0Q^2(u - 2\mu)}{(3\mu u - 2Q^2 - u^2)^{\frac{5}{2}}}\exp\left(-\frac{\arctan\left(\frac{3\mu - 2u}{\sqrt{8Q^2 - 9\mu^2}}\right)}{\sqrt{\frac{8Q^2}{9\mu^2} - 1}}\right). \tag{3.12}$$

From formulas (3.10), (3.12) it can be seen that for the functions $f(\phi(u))$, $U$ and $h(u)$ to be real, the following condition must hold

$$Q < \frac{3\mu}{2\sqrt{2}}, \tag{3.13}$$

and at the points

$$u_{*1} = \frac{3\mu}{2}\left(1 - \sqrt{1 - \frac{8Q^2}{9\mu^2}}\right) \tag{3.14}$$

and

$$u_{*2} = \frac{3\mu}{2}\left(1 + \sqrt{1 - \frac{8Q^2}{9\mu^2}}\right) \qquad (3.15)$$

these functions are not formally defined. Thus $u_{*1}$ and $u_{*2}$ are breaking points of the functions $f(\phi(u))$, $U$ and $h(u)$.

As our research shows, real functions $f(\phi(u))$, $U$ and $h(u)$ are defined only in the interval $u \in (u_{*1}, u_{*2})$ when condition (3.13) is satisfied. From a physical point of view, the functions $f(\phi(u))$ and $h(u)$ must always be positive. If we take this condition into account, then the domain of these functions is further reduced and becomes $u \in (u_0, u_{*2})$, where $u_0 = 2\mu$. As can be seen from (3.12), the function $h(u)$ vanishes at the point $u_0 = 2\mu$.

## 3.2 Bocharova-Bronnikov-Melnikov-Bekenstein-(Anti)de-Sitter metric

In the extremal case, when $\mu = Q$, we obtain a new metric from Reissner-Nordström-(Anti-)de Sitter metric. This metric can be called the Bocharova-Bronnikov-Melnikov-Bekenstein-(Anti)de-Sitter (BBMB-(A)dS) metric and presented in the following form:

$$ds^2 = -\left(\left(1 - \frac{\mu}{u}\right)^2 - \frac{\Lambda}{3}u^2\right)dt^2 + \frac{1}{\left(1 - \frac{\mu}{u}\right)^2 - \frac{\Lambda}{3}u^2}du^2 + u^2 d\Omega^2. \qquad (3.16)$$

As it is seen above, the lapse function $A(u)$ and central function $C(u)$ have form:

$$A(u) = \left(1 - \frac{\mu}{u}\right)^2 - \frac{\Lambda}{3}u^2, \qquad C(u) = u^2. \qquad (3.17)$$

If $\Lambda = 0$, Bocharova, Bronnikov, Melnikov [10] and Bekenstein [11] studied the Black hole solutions of the action.

As we know, the astrophysical parameters and properties of the Bocharova-Bronnikov-Melnikov-Bekenstein-(Anti)de-Sitter (BBMB-(A)dS) black hole have not yet been fully studied in theoretical research. We have managed to calculate the some astrophysical parameters of the black hole in this metric. The event horizon $p_h$ is the most important and fascinating characteristic of a black hole. Simply put, it is the boundary, the "point of no return." Once an object (whether a star, a speck of dust, or even light) crosses this boundary, it can never go back and will inevitably fall into the center of the black hole (the singularity). As shown in our calculations, there are three positive horizons in case of $\Lambda > 0$:

$$u_{h1} = \frac{1 - \sqrt{1 - 4\mu\sqrt{\frac{\Lambda}{3}}}}{2\sqrt{\frac{\Lambda}{3}}} \qquad - \text{Inner horizon}, \qquad (3.18)$$

$$u_{h2} = \frac{1 + \sqrt{1 - 4\mu\sqrt{\frac{\Lambda}{3}}}}{2\sqrt{\frac{\Lambda}{3}}} \quad \text{– Event horizon.} \tag{3.19}$$

Two positive roots if $\Lambda < \frac{3}{16\mu^2}$ (non-extremal).

$$u_{h3} = \frac{\sqrt{1 + 4\mu\sqrt{\frac{\Lambda}{3}}} - 1}{2\sqrt{\frac{\Lambda}{3}}} \quad \text{– Cosmological horizon.} \tag{3.20}$$

In case of $\Lambda < 0$, there is no any horizon (naked singularity). The case $\Lambda = 0$ is considered as a extremal one, and we find a single horizon $u_h = \mu$.

The innermost stable circular orbit (ISCO) $u_c$ and the photon sphere $u_{ph}$ are two distinct, critical radii around a black hole. They define the limits of orbital mechanics in General Relativity. The key distinction is that the ISCO is for massive particles (the last stable orbit), while the photon sphere is for massless particles (the last circular orbit, always unstable).

Photon sphere of the BBMB-(A)dS black hole equal to $u_{ph} = 2\mu$. The innermost stable circular orbit (ISCO) is found by solving next equaton:

$$A(u_c)\ddot{A}(u_c) - 2(\dot{A}(u_c))^2 + \frac{3A(u_c)\dot{A}(u_c)}{u_c} = 0. \tag{3.21}$$

In the special case, when $\Lambda = 0$, the innermost stable circular orbit (ISCO) is $u_{c1} = 4\mu$. In case of $\Lambda > 0$ (de-Sitter), the ISCO radius decreases compared to $4\mu$ ($u_c < 4\mu$), and otherwise, when $\Lambda < 0$ (anti-de-Sitter), the ISCO radius increases compared to $4\mu$ ($u_c > 4\mu$).

Now, let us consider application of our results obtained on the base of reconstructed scalar-tensor theory with nonminimal coupling to the BBMB-(A)dS metric. For the functions $F$ and $G$ from (2.30), (2.31), respectively, we obtain

$$F = -\frac{2(u-\mu)(u-2\mu)}{u}, \tag{3.22}$$

$$G = \left(\frac{2\mu}{u}\right)^2, \tag{3.23}$$

and hence the master equation (2.29) has solution:

$$f = -C_0\left(1 - \left(\frac{u}{u-\mu}\right)^2\right). \tag{3.24}$$

The calculation of the functions $E_U$, $F_U$ and $G_U$ from (2.25), (2.26) and (2.27) gives us

$$E_U = -\frac{1}{2}\left[\left(1 - \frac{\mu}{u}\right)^2 - \frac{\Lambda}{3}u^2\right], \tag{3.25}$$

$$F_U = \Lambda u - \frac{\mu^2}{u^3} + \frac{3\mu}{u^2} - \frac{2}{u}, \tag{3.26}$$

$$G_U = \Lambda + \left(\frac{\mu}{u^2}\right)^2 \tag{3.27}$$

and hence we find from (2.24)

$$U = -\frac{C_0 \Lambda u\left(u - 2\mu\right)\left(\left(u - \mu\right)^2 + \mu^2\right)}{\left(u - \mu\right)^4}. \tag{3.28}$$

It is seen from the last formula, at the big enough value of $u$:

$$U \sim -C_0 \Lambda.$$

From relation (2.28) we get

$$h(u) = \dot{\phi}^2 = -\frac{6C_0\mu^2}{(u-\mu)^4}, \tag{3.29}$$

and solution of this equaton is

$$\phi - \phi_0 = \pm\sqrt{-6C_0}\left(\frac{\mu}{u-\mu}\right), \tag{3.30}$$

where $\phi_0$ is a constant. From (3.30) we see, that for a canonical scalar field solution must be $C_0 < 0$. Reverting this relation we get

$$u = \pm\left(\frac{\mu\sqrt{-6C_0}}{\phi - \phi_0}\right) + \mu. \tag{3.31}$$

It follows from (3.24) and (3.31) that

$$f(\phi) = -\left(\frac{(\phi - \phi_0)^2}{6} \pm \sqrt{-\frac{2C_0}{3}}(\phi - \phi_0)\right). \tag{3.32}$$

As required, the function must always be positive, so the coupling function $f(\phi)$ has the following final form:

$$f(\phi) = \sqrt{-\frac{2C_0}{3}}(\phi - \phi_0) - \frac{(\phi - \phi_0)^2}{6} > 0, \tag{3.33}$$

where a canonical scalar field $\phi$ is defined in the domain $\phi \in (\phi_0, \phi_0 + 2\sqrt{-6C_0})$, i. e. the interval of the domain of the function $\phi(u)$ depends on the negative constant $C_0$.

# 4 Conclusion

The relevance of scalar-tensor theories has been rejuvenated in the past quarter-century by observational cosmology. The discovery that the universe's expansion is accelerating requires either a mysterious "dark energy" component within GR or a modification of gravity itself on the largest scales . Scalar-tensor theories provide a natural framework for such modifications.

In this paper we considered the scalar-tensor theory with nonminimal coupling in the Jordan frame. The action of the model contains potential term $U(\phi)$, a coupling function $f(\phi)$.

We have explored a reconstruction procedure for a generic static spherically symmetric metric written in the Buchdahl parametrization of the radial coordinate $u$: $ds^2 = (A(u))^{-1} du^2 - A(u)dt^2 + C(u)d\Omega^2$, with given $A(u) > 0$ and $C(u) > 0$. Similarly to how it was done in Ref. [12] we have found relations for $U(\phi(u))$, $f(\phi(u))$ and $\frac{d\phi}{du}$, which give us exact solutions to equations of motion with a given metric. The solutions under consideration are defined up to solutions of master equation, which is a first-order linear differential equation for the function $f(\phi(u))$. We have shown that the solutions to master equation exist for all $A(u) > 0$ and $C(u) > 0$.

In this paper we consider the scalar-tensor gravity theory with nonminimal coupling, focusing on the reconstruction method. The application of the obtained results based on the reconstruction method is demonstrated using examples in the Reissner-Nordström-(Anti-)de Sitter- and Bocharova-Bronnikov-Melnikov-Bekenstein-(Anti)de-Sitter metrics. In the result of our research we find the potential $U$ in explicit form, the domain of definition of the canonical scalar field $\phi(u)$, and the coupling function $f(u)$. In addition, we have explicitly defined the coupling functions $f(\phi)$ and $h(u)$ in the given metrics. And besides, some main astrophysical parameters of the Bocharova-Bronnikov-Melnikov-Bekenstein-(Anti)de-Sitter black hole are found. In the BBMB-(A)dS metric, a canonical scalar field $\phi(u)$ is defined in explicit form

We anticipate that the proposed reconstruction formalism for a static, spherically symmetric metric (expressed in terms of the Buchdahl parametrization) can be extended to certain generalizations of the scalar-tensor gravity theory, such as those incorporating a perfect fluid [13] or nonlinear electrodynamics [14], [15] etc.

# Acknowledgements

The author is grateful to prof. V.D. Ivashchuk for many invaluable discussions and comments on a draft.

# Appendix

## A The Lagrangian

Here we derive the formula

$$\sqrt{-g}\left(\frac{f(\phi)R(g)}{2\kappa^2} - \frac{1}{2}g^{MN}\partial_M\varphi\partial_N\varphi - U(\varphi)\right) = L + \frac{dF_*}{du}, \tag{A.1}$$

for the metric
$$ds^2 = g_{MN}dz^M dz^N = e^{2\gamma(u)}du^2 - e^{2\alpha(u)}dt^2 + e^{2\beta(u)}d\Omega^2, \tag{A.2}$$
where
$$L = \frac{f(\phi)}{\kappa^2}\left(e^{\alpha-\gamma+2\beta}\dot{\beta}\left(\dot{\beta}+2\dot{\alpha}\right) + e^{\alpha+\gamma}\right) + \frac{1}{\kappa^2}\dot{\phi}\frac{df}{d\phi}\left(\dot{\alpha}+2\dot{\beta}\right)e^{\alpha-\gamma+2\beta} -$$
$$\frac{1}{2}e^{\alpha-\gamma+2\beta}\dot{\phi}^2 - e^{\alpha+\gamma+2\beta}U(\phi) + \frac{dF_*}{du}, \tag{A.3}$$
and
$$F_* = -\frac{f(\phi)}{\kappa^2}\left(\dot{\alpha}+2\dot{\beta}\right)e^{\alpha-\gamma+2\beta}. \tag{A.4}$$

Indeed, the scalar curvature for the metric (A.2) reads
$$R(g) = 2e^{-2\beta} - 2e^{-2\gamma}\left(\dot{\alpha}^2 + 2\dot{\alpha}\dot{\beta} + 3\dot{\beta}^2 - \left(\dot{\alpha}+2\dot{\beta}\right)\dot{\gamma} + \ddot{\alpha} + 2\ddot{\beta}\right), \tag{A.5}$$
which after multiplication by $f(\phi)|g|^{\frac{1}{2}} = f(u)e^{\alpha+\gamma+2\beta}$ gives
$$f(\phi)R(g)|g|^{\frac{1}{2}} = f(\phi)\left(2e^{\alpha+\gamma} - 2e^{\alpha+2\beta-\gamma}\left(\dot{\alpha}^2 + 2\dot{\alpha}\dot{\beta} + 3\dot{\beta}^2 - \left(\dot{\alpha}+2\dot{\beta}\right)\dot{\gamma} + \ddot{\alpha} + 2\ddot{\beta}\right)\right). \tag{A.6}$$

Using $|g|^{\frac{1}{2}} = e^{\alpha+\gamma+2\beta}$, we get
$$\frac{f(\phi)R(g)}{2\kappa^2}|g|^{\frac{1}{2}} = \frac{f(\phi)}{\kappa^2}\left(e^{\alpha-\gamma+2\beta}\dot{\beta}\left(\dot{\beta}+2\dot{\alpha}\right) + e^{\alpha+\gamma}\right) + \frac{1}{\kappa^2}\dot{\phi}\frac{df}{d\phi}\left(\dot{\alpha}+2\dot{\beta}\right)e^{\alpha-\gamma+2\beta} + \frac{dF_*}{du}, \tag{A.7}$$
where
$$F_* = -\frac{f(\phi)}{\kappa^2}\left(\dot{\alpha}+2\dot{\beta}\right)e^{\alpha-\gamma+2\beta}. \tag{A.8}$$

Using $g^{MN}\partial_M\varphi\partial_N\phi = g^{uu}\dot{\phi}^2 = e^{-2\gamma}\dot{\phi}^2$ and formula (A.7) we get relations (A.1), (A.3), (A.4).

# References

[1] P.G. Bergmann, Comments on the scalar-tensor theory. Int. J. Theor. Phys., 1, (1968), https://doi.org/10.1007/BF00668828,

[2] R. V. Wagoner, Scalar-Tensor Theory and Gravitational Waves, Phys. Rev. D 1, 3209(1970), https://doi.org/10.1103/PhysRevD.1.3209,

[3] B.M. Barker, General scalar-tensor theory of gravity with constant G, Astrophys. J. 219, 5 (1978).

[4] J. Schwinger, Particles, Sources and Fields (Addison-Wesley, Reading, MA, Vol. 1, 1970).

[5] [30] William Bruckman, Generation of electro and magneto static solutions of the scalar-tensor theories of gravity, arXiv: gr-qc/9407003.

[6] K.A. Bronnikov, S. V. Bolokhov, M. Skvortsova, R. Ibadov, F. Shaymanova, On the stability of electrovacuum space-times in scalar–tensor gravity, The European Physical Journal C, 84, 1027(2024), https://doi.org/10.1140/epjc/s10052-024-13420-2

[7] K.A. Bronnikov, K. Badalov, R. Ibadov, Arbitrary Static, Spherically Symmetric Space-Times as Solutions of Scalar-Tensor Gravity. Gravit. Cosmol. 29, (2023). https://doi.org/10.1134/S0202289323010036

[8] L. Perivolaropoulos, F. Skara, Scalar tachyonic instabilities in gravitational backgrounds: Existence and growth rate. Phys. Rev. D 102(2020), 104034, https://doi.org/10.1103/PhysRevD.102.104034

[9] V. K. Oikonomou, Reissner–Nordström Anti-de Sitter Black Holes in Mimetic F(R) Gravity. Universe 2016, 2, 10. https://doi.org/10.3390/universe2020010

[10] N.M. Bocharova, K.A. Bronnikov and V.N. Melnikov, On an exact solution of the system of Einstein equations and massless scalar field, Vestn. Mosk. Univ., Fiz. Astron. 1970, No. 6, 706–709.

[11] J. D. Bekenstein, Exact solutions of Einstein conformal scalar equations, Annals Phys. 82 (1974) 535–547.

[12] K. K. Ernazarov, V. D. Ivashchuk, The problem of reconstruction for static spherically-symmetric $4d$ metrics in scalar-Einstein-Gauss-Bonnet model. The European Physical Journal C, 85: 756 (2025), [arXiv:2503.00244 [gr-qc]]

[13] V. Faraoni, S. Giardino, A. Giusti, R. Vanderwee, Scalar field as a perfect fluid: thermodynamics of minimally coupled scalars and Einstein frame scalar-tensor gravity, The European Physical Journal C 83, 24(2023). https://doi.org/10.1140/epjc/s10052-023-11186-7

[14] G. Alencar, K.A. Bronnikov, M.E. Rodrigues et al. On black bounce space-times in non-linear electrodynamics. Eur. Phys. J. C 84, 745(2024). https://doi.org/10.1140/epjc/s10052-024-13119-4

[15] I. Zh. Stefanov, S. S. Yazadjiev, M. D. Todorov, Scalar-tensor black holes coupled to Born-Infeld nonlinear electrodynamics, Phys. Rev. D 75, 084036(2007). https://doi.org/10.1103/PhysRevD.75.084036